\documentclass[
  reprint,
  aps,
  prx,
  superscriptaddress,
  longbibliography,
  nobalancelastpage
]{revtex4-2}

\usepackage{newtxtext}
\usepackage{newtxmath}
\usepackage{graphicx}
\usepackage{bm}
\usepackage{microtype}

\usepackage[colorlinks=true,citecolor=blue,urlcolor=blue,linkcolor=black]{hyperref}

\makeatletter
\newif\ifrfci@floatallowed
\newcommand{\FigurePageColumn}[2]{%
  \expandafter\xdef\csname rfci@page@\number\@currbox\endcsname{#1}%
  \expandafter\xdef\csname rfci@column@\number\@currbox\endcsname{#2}%
}
\def\rfci@checkfloat{%
  \rfci@floatallowedtrue
  \clearpage@sw{}{%
    \ifcsname rfci@page@\number\@currbox\endcsname
      \edef\rfci@targetpage{\csname rfci@page@\number\@currbox\endcsname}%
      \edef\rfci@targetcolumn{\csname rfci@column@\number\@currbox\endcsname}%
      \ifnum\rfci@targetpage>0\relax
        \ifnum\c@page<\rfci@targetpage\relax
          \rfci@floatallowedfalse
        \else
          \ifnum\c@page=\rfci@targetpage\relax
            \ifnum\rfci@targetcolumn>0\relax
              \ifnum\pagegrid@cur=\rfci@targetcolumn\relax\else
                \rfci@floatallowedfalse
              \fi
            \fi
          \fi
        \fi
      \fi
    \fi
  }%
  \ifrfci@floatallowed\else
    \global\let\@deferlist@postshipout\@empty
  \fi
}
\AtBeginDocument{%
  \let\rfci@original@xfloat\@xfloat
  \def\@xfloat#1[#2]{%
    \rfci@original@xfloat{#1}[#2]%
    \expandafter\xdef\csname rfci@page@\number\@currbox\endcsname{0}%
    \expandafter\xdef\csname rfci@column@\number\@currbox\endcsname{0}%
  }%
  \let\rfci@original@addtocurcol\@addtocurcol
  \def\@addtocurcol{%
    \rfci@checkfloat
    \ifrfci@floatallowed
      \rfci@original@addtocurcol
    \else
      \@insertfalse
      \@resethfps
      \@cons\@deferlist\@currbox
    \fi
  }%
  \let\rfci@original@addtonextcol\@addtonextcol
  \def\@addtonextcol{%
    \rfci@checkfloat
    \ifrfci@floatallowed
      \rfci@original@addtonextcol
    \else
      \@insertfalse
      \@cons\@deferlist\@currbox
    \fi
  }%
  \let\rfci@original@addtodblcol\@addtodblcol
  \def\@addtodblcol{%
    \rfci@checkfloat
    \ifrfci@floatallowed
      \ifnum\c@page=3\relax
        \begingroup
          \let\rfci@saved@deferlist\@deferlist
          \global\let\@deferlist\@empty
          \rfci@original@addtodblcol
          \let\@elt\relax
          \xdef\@deferlist{\rfci@saved@deferlist\@deferlist}%
        \endgroup
      \else
        \rfci@original@addtodblcol
      \fi
    \else
      \@cons\@deferlist\@currbox
    \fi
  }%
  \let\rfci@original@combinedblfloats\@combinedblfloats
  \def\@combinedblfloats{%
    \ifnum\c@page=3\relax
      \@ifx@empty\@dbltoplist{}{%
        \setbox\@tempboxa\vbox{}%
        \let\@elt\@comdblflelt\@dbltoplist
        \let\@elt\relax
        \xdef\@freelist{\@freelist\@dbltoplist}%
        \global\let\@dbltoplist\@empty
        \setbox\@outputbox\vbox{%
          \unvbox\@outputbox
          \vskip\dbltextfloatsep
          \unvbox\@tempboxa\unskip
        }%
      }%
    \else
      \rfci@original@combinedblfloats
    \fi
  }%
}
\newcommand{\RestoreDefaultFigurePlacement}{%
  \let\@xfloat\rfci@original@xfloat
  \let\@addtocurcol\rfci@original@addtocurcol
  \let\@addtonextcol\rfci@original@addtonextcol
  \let\@addtodblcol\rfci@original@addtodblcol
  \let\@combinedblfloats\rfci@original@combinedblfloats
}
\makeatother

\newcommand{\figplaceholder}[2]{%
  \fbox{%
    \parbox[c][#1][c]{0.96\linewidth}{%
      \centering\sffamily\small #2%
    }%
  }%
}

\begin{document}


\title{
Fractional Chern Insulators in Laser-Addressed Dipolar Rydberg Atom Arrays
}



\author{Xingke Wang}
\affiliation{Institute of Physics, Chinese Academy of Sciences, Beijing 100190, China}
\affiliation{School of Physical Sciences, University of Chinese Academy of Sciences, Beijing 100049, China}

\author{Cheng Chen}
\email{cheng.chen@iphy.ac.cn}
\affiliation{Institute of Physics, Chinese Academy of Sciences, Beijing 100190, China}



\begin{abstract}
We predict the existence of fractional Chern insulators in laser-addressed dipolar Rydberg atom arrays.
Importantly, we propose a scheme to construct a uniform synthetic flux in a dipolar Rydberg atom array by choosing a bilayer checkerboard lattice and applying addressing lasers that bring the selected dipolar-exchange channel near resonance.
In this bilayer configuration, we find that the system hosts an isolated topological flat band with Chern number $C=1$ and a bosonic $\nu=1/2$ fractional Chern insulator.
In a single-layer coplanar configuration, the flux pattern depends on the geometric parameter $h$, generating either a uniform or a staggered flux.
The uniform-flux configuration also hosts a fractional Chern insulator.
The many-body gap remains finite along an adiabatic interpolation between the bilayer and coplanar configurations, providing evidence for adiabatic continuity.
The tunable flux configurations provide opportunities to study further topological states of matter in laser-addressed dipolar Rydberg atom arrays.

\end{abstract}

\maketitle

\begin{figure*}[t]
  \FigurePageColumn{2}{0}
  \centering
  \IfFileExists{figures/fig1.pdf}{%
    \includegraphics[width=\textwidth,height=0.60\textheight,keepaspectratio]{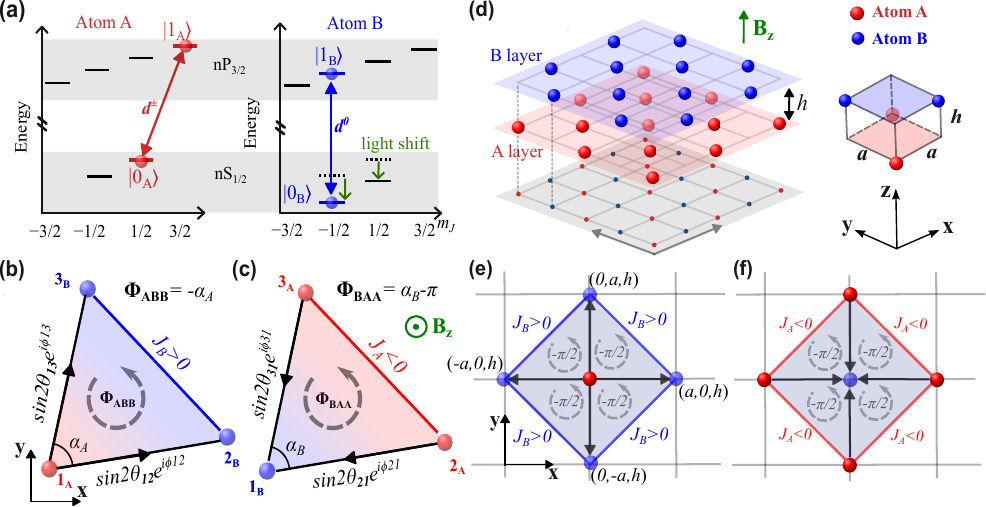}%
  }{%
    \figplaceholder{5.6cm}{fig1.pdf\\[1ex]
    Bilayer geometry, local levels, and loop fluxes}%
  }
\caption{
\textbf{Synthetic flux in the bilayer Rydberg array.}
(a) Sublattice-dependent Rydberg-state encoding and differential-light-shift tuning of the selected $A$--$B$ pair states near resonance.
(b,c) Elementary $ABB$ and $BAA$ triangular loops projected onto the $xy$ plane.
The bond angles $(\theta_{ij},\phi_{ij})$ are defined with respect to $\mathbf r_{ij}=\mathbf r_j-\mathbf r_i$, and the complex $A$--$B$ hopping amplitudes are specified for the $A\to B$ direction.
(d) Bilayer checkerboard geometry, with projected lattice spacing $a=9~\mu\mathrm{m}$, layer separation $h$, and magnetic field $B_z$ defining the quantization axis.
(e) Directed hopping pattern and (f) the resulting uniform synthetic flux pattern in the bilayer geometry.
With $J_A<0$ and $J_B>0$, the elementary $ABB$ and $BAA$ loops both carry a flux of $-\pi/2$.
}

  \label{fig:mechanism}
\end{figure*}

\begin{figure}[!t]
  \FigurePageColumn{3}{1}
  \centering
  \IfFileExists{figures/fig2.pdf}{%
    \includegraphics[width=\columnwidth]{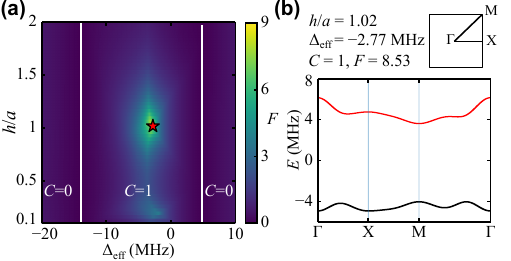}%
  }{%
    \figplaceholder{5.2cm}{fig2.pdf\\[1ex]
    Band-flatness map and single-particle spectrum}%
  }
  \caption{
\textbf{Single-particle topology and band flatness.}
(a) Lower-band flatness ratio $F=\Delta_{\rm ind}/W$ as a function of $h/a$ and $\Delta_{\rm eff}$.
All hopping amplitudes within the three-dimensional cutoff $R_c=3a$ are included.
White curves mark direct band-gap-closing boundaries separating regions with different lower-band Chern numbers, and the star marks the optimal single-particle flat-band point.
(b) Single-particle spectrum at the marked point in (a), showing an isolated lower band with $C=1$ and $F=8.53$. 
}

  \label{fig:singleparticle}
\end{figure}

\begin{figure*}[!t]
  \FigurePageColumn{3}{0}
  \centering
  \IfFileExists{figures/fig3.pdf}{%
    \includegraphics[width=\textwidth]{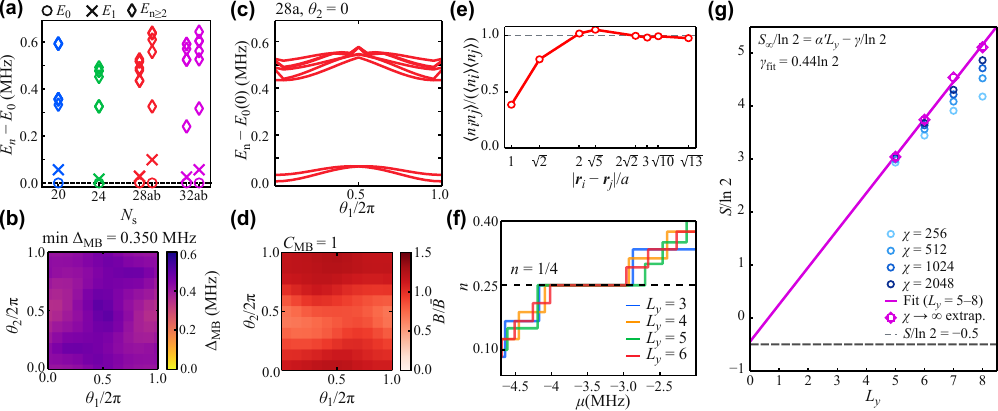}%
  }{%
    \figplaceholder{8.6cm}{fig3.pdf\\[1ex]
    (a) spectra \quad (b) twist-angle gap \\[0.8ex] (c) spectral flow \quad (d) Berry curvature \\[0.8ex] (e) incompressibility \quad (f) correlations \\[0.8ex] (g) entanglement entropy}%
  }
 \caption{
\textbf{Many-body signatures of the bosonic $\nu=1/2$ Laughlin FCI.}
(a) Zero-twist low-energy spectra for tori of different sizes and shapes.
(b) Neutral gap $\Delta_{\mathrm{MB}}(\theta_1,\theta_2)$ on an $11\times11$ twist-angle grid for $N_s=28$.
(c) Spectral flow of the lowest eight $N_s=28$ levels along $(\theta_1,0)$ for $0\leq\theta_1\leq2\pi$, shown as $E_n(\theta_1,0)-E_0(0,0)$.
(d) Berry flux of the rank-2 low-energy subspace for $N_s=28$, giving a total many-body Chern number $C_{\mathrm{MB}}=1$.
(e) Normalized density-density correlation obtained from exact diagonalization, plotted as a function of the projected separation $r$ in the $xy$ plane.
(f) Particle density as a function of chemical potential, obtained from infinite-cylinder MPS calculations for $L_y=3,4,5,6$ at bond dimension $\chi=512$, showing a plateau at $n=1/4$ within the numerical resolution.
(g) Bond-dimension-extrapolated entanglement entropy $S_\infty(L_y)$ for $L_y=5,6,7,8$, yielding $\gamma\simeq0.44\ln2$.
}

  \label{fig:fci}
\end{figure*}

\begin{figure}[!t]
  \FigurePageColumn{4}{2}
  \centering
  \includegraphics[width=\columnwidth]{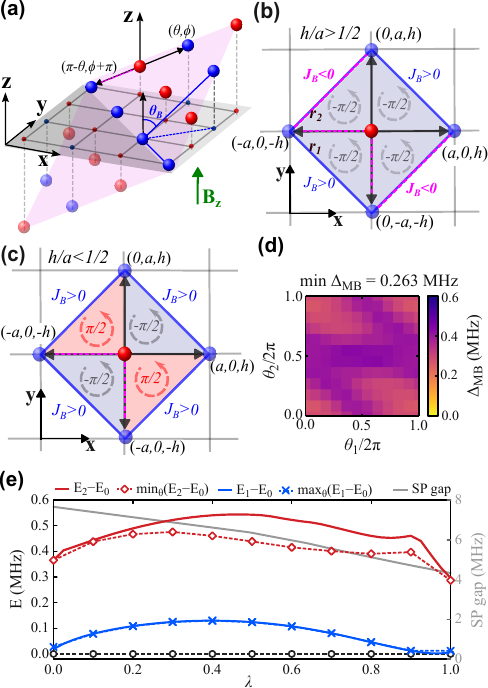}
  \caption{
\textbf{Geometry-controlled coplanar flux and adiabatic interpolation.}
(a) coplanar configuration with $\mathbf r_{n,m}=(na,ma,(n+m)h)$; sites with even and odd $n+m$ form sublattices $A$ and $B$, respectively.
(b) Uniform flux for $h/a>1/2$, with $\Phi=-\pi/2$ on every elementary triangle.
(c) Staggered flux for $0<h/a<1/2$: triangles containing horizontal and tilted intrasublattice diagonals carry fluxes $-\pi/2$ and $+\pi/2$, respectively.
Pink bonds denote hopping terms with signs reversed relative to the bilayer configuration.
(d) Neutral many-body gap $\Delta_{\mathrm{MB}}=E_2-E_1$ for the coplanar endpoint on an $11\times11$ twist-angle grid with $N_s=28$, with a minimum of $0.263~\mathrm{MHz}$.
(e) Gauge-aligned interpolation from the bilayer ($\lambda=0$) to the coplanar endpoint ($\lambda=1$). Solid red and blue curves show $E_2-E_0$ and $E_1-E_0$, respectively, at zero twist. 
Markers indicate their respective minima and maxima over a $6\times6$ twist-angle grid, as specified in the legend. The gray curve shows the single-particle indirect gap (right axis).
}

  \label{fig:coplanar}
\end{figure}


\textit{Introduction.—}
The fractional quantum Hall effect is a paradigmatic form of interaction-driven topological order, with the Laughlin state as its canonical realization~\cite{Laughlin1983}.
Fractional Chern insulators (FCIs) extend this physics to partially filled Chern bands without Landau levels~\cite{Tang2011,Neupert2011,Sheng2011,RegnaultBernevig2011}.
A favorable regime for FCIs combines a narrow, well-isolated Chern band with interactions strong compared with its residual bandwidth~\cite{Parameswaran2013,BergholtzLiu2013}.

Rydberg atom arrays provide a programmable platform for quantum simulation, combining reconfigurable geometry, strong dipolar interactions, and local control~\cite{Bernien2017,Ebadi2021,Scholl2021,Semeghini2021,ChenContinuous2023,Bornet2023,Bornet2024,ChenSpectroscopy2025,EmperaugerTLL2025,Qiao2025,Bornet2026,ChenWeakConservation2026,RuanElectronSpin2026,BaiRydbergReview2026}.
The intrinsic spin-orbit coupling of dipolar interactions introduces complex hopping terms~\cite{Yao2012,Yao2013,Syzranov2014,Peter2015,Kiffner2017,Weber2018}, generating synthetic flux and breaking time-reversal symmetry~\cite{Lienhard2020}.
However, in existing dipolar Rydberg atom array experiments, Zeeman splitting among magnetic sublevels under an external magnetic field typically suppresses phase-dependent dipolar exchange.
Recent advances in site-resolved laser addressing allow state-dependent AC Stark shifts to bring selected dipolar-exchange processes near resonance~\cite{deLeseleuc2017,Bornet2024,Emperauger2025}.
Motivated by exploring topological physics in dipolar Rydberg atom arrays, we propose that combining such addressing beams with tailored lattice geometries enables programmable synthetic flux patterns—such as uniform or staggered flux—in these arrays.
Importantly, we find supporting evidence for fractional Chern insulators within this framework, establishing a versatile route to strongly correlated topological states in laser-addressed Rydberg arrays.
%

%
Our results cover the following three aspects:
First, we propose a scheme to generate a uniform-flux pattern in dipolar Rydberg atom arrays.
In a bilayer checkerboard array, we apply site-resolved addressing beams to induce local light shifts that bring the phase-dependent dipolar-exchange channel near resonance.
The resulting atomic-orientation-dependent phases generate a uniform synthetic magnetic flux through the elementary triangular loops.
Second, we find that the resulting single-particle spectrum contains an isolated, nearly flat band with Chern number $C=1$.
Importantly, at filling $\nu=1/2$, exact diagonalization reveals a twofold low-energy manifold separated from higher states by a finite neutral gap.
We show that the spectral flow under flux insertion, quantized total many-body Chern number $C_{\mathrm{MB}}=1$, and matrix-product-state results support the identification of a bosonic Laughlin FCI.
Third, we propose a coplanar realization in which tuning the geometric parameter $h$ allows switching between uniform and staggered flux configurations, with the former also hosting a bosonic $\nu=1/2$ Laughlin FCI.
For the uniform-flux configuration, we construct an adiabatic interpolation to the bilayer case.
The single-particle and many-body gaps remain finite at all sampled points along the interpolation, providing finite-size evidence for adiabatic continuity between the two realizations.

\medskip

\textit{Setup.—}
Our scheme is based on $^{87}\mathrm{Rb}$ dipolar Rydberg atom arrays.
Using a spherical-tensor expansion, the Rydberg dipole-dipole interaction can be written as

\begin{equation}
\begin{aligned}
\hat V_{ij}^{\mathrm{dd}}
=
&\frac{1}{4\pi\epsilon_0 r_{ij}^3}
\Bigg\{
(1-3\cos^2\theta_{ij})
\left[
d_i^0d_j^0
+\frac{1}{2}
\left(d_i^+d_j^-+d_i^-d_j^+\right)
\right]
\\
&+\frac{3}{2\sqrt{2}}
\sin(2\theta_{ij})
\left[
e^{-i\phi_{ij}}
\left(d_i^+d_j^0+d_i^0d_j^+\right)
+\mathrm{H.c.}
\right]
\\
&-\frac{3}{2}\sin^2\theta_{ij}
\left[
e^{-2i\phi_{ij}}d_i^+d_j^+
+\mathrm{H.c.}
\right]
\Bigg\},
\end{aligned}
\label{eq:Vdd}
\end{equation}
where $d^0=d_z$ and $d^\pm=\mp(d_x\pm i d_y)/\sqrt{2}$, and $(r_{ij},\theta_{ij},\phi_{ij})$ specify the relative position $\mathbf r_{ij}=\mathbf r_j-\mathbf r_i$.
The three terms in Eq.~\eqref{eq:Vdd} account for processes with $\Delta M=0$, $\pm 1$, and $\pm 2$, respectively, with $M=m_{J,i}+m_{J,j}$ being the total magnetic quantum number of the pair.
Taking a two-atom system as an example, a static magnetic field $B_z\hat{\mathbf z}$ defines the quantization axis, under which the $\Delta M=\pm1$ and $\pm2$ channels are naturally off-resonance due to Zeeman splittings [Fig.~\ref{fig:mechanism}(a)].
To exploit the phase-dependent $\Delta M=\pm1$ channel, we apply site-resolved addressing beams to induce local light shifts—specifically shifting the $S$ state of atom $B$ downward—which brings the pair states $\vert{}0_A1_B\rangle$ and $\vert{}1_A0_B\rangle$ near resonance with a residual detuning $\Delta_{\mathrm{eff}}$ [Fig.~\ref{fig:mechanism}(a)]. 
Such dipolar exchange involving these levels has been experimentally benchmarked in a two-atom system~\cite{Emperauger2025}.

For the multi-atom array, we adopt the following encoding scheme:
On sublattice $A$, we use $\vert{}0_A\rangle=\vert{}60S_{1/2},m_J=+1/2\rangle$ and $\vert{}1_A\rangle=\vert{}60P_{3/2},m_J=+3/2\rangle$, whereas on sublattice $B$, $\vert{}0_B\rangle=\vert{}60S_{1/2},m_J=-1/2\rangle$ and $\vert{}1_B\rangle=\vert{}60P_{3/2},m_J=-1/2\rangle$.
For the exchange coupling $J_{ij}=\langle 0_i1_j|\hat V_{ij}^{\mathrm{dd}}|1_i0_j\rangle$, we denote the intrasublattice nearest-neighbor (NN) couplings by $J_{A}$ ($J_{B}$), and the intersublattice NN coupling by $J_{AB}$.
The intersublattice coupling $J_{AB}$ arises from the circular--longitudinal $\Delta M=\pm1$ channel, which has angular dependence $J_{AB}\propto\sin(2\theta_{AB})\,e^{i\phi_{AB}}$.
At a fixed angle $\theta$, the intrasublattice couplings $J_A$ and $J_B$ have opposite signs.
This is because they originate from the $d_i^+d_j^-+d_i^-d_j^+$ and $d_i^0d_j^0$ channels, respectively; the relative minus sign arises from $(d^+)^\dagger=-d^-$.

We now evaluate the synthetic flux through the two elementary mixed-sublattice triangles, $ABB$ and $BAA$ [Figs.~\ref{fig:mechanism}(b) and \ref{fig:mechanism}(c)].
For a counterclockwise loop $(1,2,3)$, we define the synthetic flux as $\Phi=\arg\!\left(J_{12}J_{23}J_{31}\right)$.
The loop flux is determined by the azimuthal phases $\phi_{ij}$ and the signs of $\sin(2\theta)$ in $J_{AB}$, and the signs of $J_{A}$, $J_{B}$.
For the $ABB$ triangle, when $J_{B}>0$ and $\theta_{AB} \in (0,\pi/2)$, the loop has a flux of $\Phi_{ABB}=-\alpha_A$, where $\alpha_A=\phi_{13}-\phi_{12}$.
For the $BAA$ triangle, when $J_{A}<0$ and $\theta_{AB} \in (0,\pi/2)$, the loop has a flux of $\Phi_{BAA}=\alpha_B-\pi$, where $\alpha_B=\phi_{31}-\phi_{21}$.
We consider a bilayer checkerboard array with atomic positions $\mathbf r_A=(na,ma,0)$ for even $n+m$ and $\mathbf r_B=(na,ma,h)$ for odd $n+m$, where $a=9~\mu\mathrm{m}$ is the projected lattice spacing and  $h>0$ is the layer separation [Fig.~\ref{fig:mechanism}(d)].
In this geometry, $J_A<0$, $J_B>0$, and $\sin(2\theta_{AB})>0$; together with $\alpha_A=\alpha_B=\pi/2$, these yield a uniform synthetic flux $\Phi_{ABB}=\Phi_{BAA}=-\pi/2$ through all elementary triangles [Fig.~\ref{fig:mechanism}(e,f)].

\medskip

\textit{Topological flat band.—}
With the selected Rydberg states, the dipolar exchange gives the effective hopping Hamiltonian
\begin{equation}
H_{0}=
\sum_{i<j}
\left(
t_{ij}c_j^\dagger c_i+\mathrm{H.c.}
\right)
+\sum_i\epsilon_i n_i,
\label{eq:hopping_model_hamiltonian}
\end{equation}
where $t_{ij}=J_{ij}$ denotes the hopping amplitude from site $i$ to site $j$, with $t_{ji}=t_{ij}^{*}$, $c_i^\dagger=|1_{A(B)}\rangle_i\langle0_{A(B)}|$ is the bosonic creation operator on site $i$ of sublattice $A$ ($B$), and $n_i=c_i^\dagger c_i$.
The on-site energy is $\epsilon_i=\epsilon_A$ ($\epsilon_B$) on sublattice $A$ ($B$), and the effective sublattice detuning is defined as $\Delta_{\mathrm{eff}}=\epsilon_B-\epsilon_A$.
In the subsequent discussion, the numerical calculations use coupling strengths obtained from PairInteraction~\cite{WeberPairInteraction2017}, with interactions truncated at the three-dimensional distance $R_c=3a$.

We now focus on the single-particle band structure which has two bands~\cite{SM}.
%
%
The synthetic flux breaks time-reversal symmetry, allowing the bands to have a nonzero Chern number~\cite{Haldane1988}.
As shown in Figure~\ref{fig:singleparticle}(a), we find a region where the lower band carries Chern number $C=1$, by tuning the layer separation $h$ and the residual detuning $\Delta_{\mathrm{eff}}$.
The color scale shows the lower-band flatness $F=\Delta_{\mathrm{ind}}/W$~\cite{Sun2011,BergholtzLiu2013}, where $W$ is the bandwidth and $\Delta_{\mathrm{ind}}$ is the band gap.
%
%
We obtain the optimal flatness $F=8.53$ at $h/a=1.02$ and $\Delta_{\mathrm{eff}}=-2.77~\mathrm{MHz}$, with $W=0.90~\mathrm{MHz}$ and $\Delta_{\mathrm{ind}}=7.66~\mathrm{MHz}$ [Fig.~\ref{fig:singleparticle}(b)].
This nearly flat topological band provides favorable conditions for a bosonic fractional Chern insulator.

\medskip

\textit{Fractional Chern insulator.—}
We now investigate whether the half-filled Chern band supports a bosonic $\nu=1/2$ Laughlin state.
Including the van der Waals (vdW) interaction, the full Hamiltonian of the hard-core bosons reads
\begin{equation}
H=H_0+\sum_{i<j}V_{ij}n_i n_j,
\label{eq:many_body_hamiltonian}
\end{equation}
where $V_{ij}$ denotes the vdW density-density interaction between sites $i$ and $j$.
The bosons satisfy the hard-core constraint $(c_i^\dagger)^2=0$.
With each unit cell containing two sites, half filling of the lower band ($\nu=1/2$) corresponds to an average site occupation of $n=\nu/2=1/4$.

At the optimal flat-band point identified above, we examine the low-energy spectrum using exact diagonalization (ED) on tori of different sizes and shapes~\cite{SM}.
We find a nearly degenerate ground-state doublet separated from higher excitation states, consistent with the twofold ground-state degeneracy of a bosonic $\nu=1/2$ Laughlin state~\cite{Wang2011,Weber2022} [Fig.~\ref{fig:fci}(a)].
For the $N_s=28$ torus (geometry 28a in Fig.~\ref{fig:torus_shapes}), the doublet splitting at zero twist is $\delta=E_1-E_0=0.0277~\mathrm{MHz}$, much smaller than the neutral many-body gap $\Delta_{\mathrm{MB}}=E_2-E_1=0.407~\mathrm{MHz}$.
To examine the topological response of this doublet, we impose twisted boundary conditions~\cite{NiuThoulessWu1985} with twist angles $\theta_1$ and $\theta_2$~\cite{SM}.
The many-body gap remains open throughout the twist-angle space, with a minimum of $0.350~\mathrm{MHz}$ [Fig.~\ref{fig:fci}(b)].
Specifically, for $\theta_1:0\to2\pi$ at $\theta_2=0$, the two low-energy states exchange while remaining separated from higher excitations [Fig.~\ref{fig:fci}(c)].
Across twist angles, the quasi-degenerate ground state exhibits a nearly uniform normalized Berry curvature $B/\bar{B}$ ($\bar{B}$ is its angle average) with a total many-body Chern number $C_{\mathrm{MB}}=1$ [Fig.~\ref{fig:fci}(d)]~\cite{Hatsugai2005,FukuiHatsugaiSuzuki2005}.
We also calculate the density-density correlations $\langle n_i n_j\rangle/
(\langle n_i\rangle\langle n_j\rangle)$ using ED to examine possible density order. 
The correlations approach unity at large separations, indicating the absence of long-range density order [Fig.~\ref{fig:fci}(e)].
We further study the ground state using infinite DMRG on cylinders that are infinite along $\mathbf a_1=a(1,-1)$ and periodic along $\mathbf a_2=a(1,1)$, with a circumference of $L_y$ unit cells~\cite{White1992,Schollwock2011,HauschildTeNPy2024}.
The particle density remains at $n=1/4$ over a finite range of chemical potentials, indicating an incompressible state [Fig.~\ref{fig:fci}(f)].
We calculate the entanglement entropy $S$ between the two halves of the infinite cylinder for different circumferences $L_y$ and bond dimensions $\chi$.
For each $L_y$, we extrapolate $S$ to $\chi\to\infty$ and obtain $S_\infty(L_y)$.
Fitting these extrapolated values to $S_\infty(L_y)=\alpha L_y-\gamma$~\cite{KitaevPreskill2006,LevinWen2006} gives a topological entanglement entropy $\gamma\simeq0.44\ln2$, close to the expected Laughlin value $\frac12\ln2$ [Fig.~\ref{fig:fci}(g)].
Taken together, these numerical results support the identification of a bosonic $\nu=1/2$ Laughlin FCI.

\medskip
\textit{Coplanar configuration and adiabatic interpolation.—}
In addition to the bilayer geometry, we investigate a coplanar configuration that is more accessible experimentally~\cite{ChenContinuous2023}.
It is tilted relative to the $xy$ plane, with atomic positions $\mathbf r_{n,m}=(na,ma,(n+m)h)$, where sites with even and odd $n+m$ form sublattices $A$ and $B$, respectively [Fig.~\ref{fig:coplanar}(a)].
The bilayer and coplanar configurations have the same square-lattice projection in the $xy$ plane, leaving the azimuthal angles $\phi_{ij}$ unchanged.
Crucially, the changes in polar angles $\theta_{ij}$ can reverse the signs of certain hopping amplitudes, thereby changing the synthetic flux pattern.

(a) For the intersublattice hopping $J_{AB}$, the displacement vectors from a site on sublattice $A$ to its four NN sites on sublattice $B$ are $\mathbf r_{AB}^{\pm x}=(\pm a,0,\pm h)$ and $\mathbf r_{AB}^{\pm y}=(0,\pm a,\pm h)$.
The bonds along $+x$ and $-x$, or $+y$ and $-y$, have polar angles $\theta$ and $\pi-\theta$, respectively.
Since both $\sin(2\theta)$ and $e^{i\phi}$ change sign between opposite NN directions, the hopping amplitudes satisfy $J_{AB}^{+x}=J_{AB}^{-x}$ and $J_{AB}^{+y}=J_{AB}^{-y}$.
By contrast, NN hoppings along opposite directions have opposite signs in the bilayer geometry.

(b) For the intrasublattice hoppings $J_A$ and $J_B$, the horizontal and tilted diagonal bonds have displacement vectors $\mathbf r_1=(a,-a,0)$ and $\mathbf r_2=(a,a,2h)$, respectively.
For $\mathbf r_1$, the polar angle is $\theta_B=\pi/2$, giving $J_B>0$.
For $\mathbf r_2$, the polar angle $\theta_B$ depends on $h$ and reaches the dipolar magic angle at $h/a=1/2$, where $J_B=0$.
Consequently, $J_B>0$ for $h/a<1/2$ and $J_B<0$ for $h/a>1/2$.
The corresponding $A$--$A$ hoppings have the opposite signs.

Consequently, for $h/a>1/2$, all elementary triangles carry a uniform synthetic flux $\Phi=-\pi/2$ [Fig.~\ref{fig:coplanar}(b)].
For $0<h/a<1/2$, triangles containing horizontal and tilted intrasublattice bonds carry fluxes $-\pi/2$ and $+\pi/2$, respectively, forming a staggered flux pattern [Fig.~\ref{fig:coplanar}(c)].

We find that the uniform-flux coplanar configuration can also support a bosonic $\nu=1/2$ Laughlin FCI.
At $h/a=1.32$ and $\Delta_{\mathrm{eff}}=-0.98~\mathrm{MHz}$, the lower single-particle band carries Chern number $C=1$, with a flatness ratio $F=3.27$ ($W=1.32~\mathrm{MHz}$ and $\Delta_{\mathrm{ind}}=4.33~\mathrm{MHz}$).
At quarter filling, ED on the $N_s=28$ torus reveals a nearly degenerate ground-state doublet, with a splitting $\delta=0.0067~\mathrm{MHz}$ and a many-body gap $\Delta_{\mathrm{MB}}=0.289~\mathrm{MHz}$ at zero twist.
The many-body gap remains open over the twist-angle space, with a minimum of $0.263~\mathrm{MHz}$ [Fig.~\ref{fig:coplanar}(d)].
The two-state manifold carries a total many-body Chern number $C_{\mathrm{MB}}=1$.

To further examine the FCI characteristic features of the coplanar system and its relation to the bilayer realization, we consider an adiabatic interpolation between the two cases~\cite{ScaffidiMoller2012,LiuBergholtz2013,Wu2012}.
Although the bilayer and coplanar configurations share the same uniform triangular flux, their hopping phases differ in the original gauges.
We therefore apply a local gauge transformation to the bilayer Hamiltonian to match the NN hopping phase of the coplanar Hamiltonian~\cite{SM}.
Using the gauge-transformed bilayer Hamiltonian $\widetilde H_{\mathrm{bi}}$, we define the linear interpolation
\begin{equation}
H(\lambda)
=
(1-\lambda)\widetilde H_{\mathrm{bi}}
+
\lambda H_{\mathrm{cop}},
\qquad 0\leq\lambda\leq1,
\label{eq:interpolation}
\end{equation}
Both Hamiltonians are defined on the same $N_s=28$ torus.
The single-particle indirect gap remains open throughout the interpolation, with a minimum of $4.33~\mathrm{MHz}$.
At $\nu=1/2$, we track $E_1-E_0$ and
$E_2-E_0$ throughout the interpolation,
both at zero twist and through their extrema
over twist-angle space
[Fig.~\ref{fig:coplanar}(e)].
The many-body gap remains open, with a minimum of $0.263~\mathrm{MHz}$ at the coplanar endpoint [Fig.~\ref{fig:coplanar}(d)].
These results support an adiabatic connection between the bilayer and coplanar realizations of the bosonic $\nu=1/2$ Laughlin FCI.

\medskip
\textit{Conclusion and outlook.—}
In summary, we propose a laser-addressed scheme for synthetic flux generation in dipolar Rydberg atom arrays, enabling the investigation of time-reversal-symmetry-breaking physics.
Specifically, in the bilayer checkerboard geometry, the site-resolved laser-induced light shifts bring the phase-dependent dipolar-exchange channels near resonance, generating a uniform synthetic flux through the elementary triangles. This gives rise to a nearly flat $C=1$ band, and our exact diagonalization, DMRG, and entanglement entropy analyses support the identification of a bosonic $\nu=1/2$ Laughlin FCI.
Furthermore, we investigate a more experimentally accessible coplanar realization, where tuning the geometric parameter $h$ allows the elementary-loop flux pattern to be precisely controlled and switched between uniform and staggered configurations.
We demonstrate that the uniform-flux coplanar configuration also hosts a robust bosonic $\nu=1/2$ Laughlin FCI, and the opening of both single-particle and many-body gaps along the gauge-aligned adiabatic interpolation provides evidence of adiabatic continuity between the bilayer and coplanar limits.
More broadly, our flux-generation scheme demonstrates that combining laser addressing with geometric design offers tunable synthetic flux, complementing existing Rydberg FCI proposals~\cite{Weber2022,Eix2025,ZhaoShi2023,Poon2024}, while relying solely on currently accessible experimental techniques.
This opens up opportunities to study a rich variety of topological states of matter, non-equilibrium quantum dynamics, and exotic many-body phenomena in programmable neutral-atom platforms.


\medskip
\textit{Acknowledgments.—}
We acknowledge Fan Yang and Yantao Wu for fruitful discussions. 
This work is supported by Quantum Science and Technology-National Science and Technology Major Project (2024ZD0301700), and the start-up grant from IOP-CAS.


\bibliography{references}


\clearpage
\RestoreDefaultFigurePlacement
\onecolumngrid
\raggedbottom

\setcounter{page}{1}
\setcounter{section}{0}
\setcounter{subsection}{0}
\setcounter{equation}{0}
\setcounter{figure}{0}
\setcounter{table}{0}
\setcounter{secnumdepth}{2}
\renewcommand{\thepage}{S\arabic{page}}
\renewcommand{\thesection}{\Roman{section}}
\renewcommand{\thesubsection}{\Alph{subsection}}
\renewcommand{\theequation}{S\arabic{equation}}
\renewcommand{\thefigure}{S\arabic{figure}}
\renewcommand{\thetable}{S\arabic{table}}
\renewcommand{\theHsection}{supp.\arabic{section}}
\renewcommand{\theHsubsection}{supp.\arabic{section}.\arabic{subsection}}
\renewcommand{\theHequation}{supp.\arabic{equation}}
\renewcommand{\theHfigure}{supp.\arabic{figure}}
\renewcommand{\theHtable}{supp.\arabic{table}}
\pagestyle{plain}
\raggedbottom
\begin{center}
{\large\bfseries Supplemental Material for\\[0.5em]
Fractional Chern Insulators in Laser-Addressed Dipolar Rydberg Atom Arrays}
\end{center}
\medskip
\twocolumngrid

\section{Single-particle Hamiltonian}
\label{sec:sm_single_particle}

To elucidate the band topology of the bilayer checkerboard array, we first consider a simplified nearest-neighbor (NN) model that includes only intersublattice and intrasublattice NN hoppings.
We define the bosonic creation operators on sublattices $A$ and $B$ as $a_i^\dagger=|1_A\rangle_i\langle0_A|$ and $b_j^\dagger=|1_B\rangle_j\langle0_B|$, respectively.
In the single-particle sector, the Hamiltonian can be written as
\begin{equation}
H_{\mathrm{NN}}=H_{AB}+H_{AA}+H_{BB}+H_{\mathrm{onsite}},
\end{equation}
with
\begin{equation}
\begin{aligned}
H_{AB}&=\sum_{\langle i,j\rangle_{AB}}\left(J_{ij}b_j^\dagger a_i+\mathrm{H.c.}\right),\\
H_{AA}&=J_A\sum_{\langle i,j\rangle_{AA}}\left(a_j^\dagger a_i+\mathrm{H.c.}\right),\\
H_{BB}&=J_B\sum_{\langle i,j\rangle_{BB}}\left(b_j^\dagger b_i+\mathrm{H.c.}\right),\\
H_{\mathrm{onsite}}&=\epsilon_A\sum_{i\in A}a_i^\dagger a_i+\epsilon_B\sum_{j\in B}b_j^\dagger b_j.
\end{aligned}
\end{equation}
Here, $H_{AB}$ describes the intersublattice hopping, while $H_{AA}$ and $H_{BB}$ describe the intrasublattice hopping on sublattices $A$ and $B$, respectively.
$\langle i,j\rangle_{AB}$ denotes intersublattice NN pairs, while $\langle i,j\rangle_{AA}$ and $\langle i,j\rangle_{BB}$ denote NN pairs within sublattices $A$ and $B$, respectively.
$H_{\mathrm{onsite}}$ describes the on-site energies, with $\epsilon_A=-\Delta_{\mathrm{eff}}/2$ and $\epsilon_B=\Delta_{\mathrm{eff}}/2$.
In momentum space, with $a_{\mathbf{k}}=N_c^{-1/2}\sum_{i\in A}a_i e^{i\mathbf{k}\cdot\mathbf r_i/a}$ and $b_{\mathbf{k}}=N_c^{-1/2}\sum_{j\in B}b_j e^{i\mathbf{k}\cdot\mathbf r_j/a}$, the Bloch Hamiltonian takes the form
\begin{equation}
\begin{aligned}
H_{\mathrm{sp}}(\mathbf k)
={}&d_0(\mathbf k)\mathbb I
+2|J_{AB}|(\sin k_x\,\sigma_y-\sin k_y\,\sigma_x)\\
&+\left[
2(J_A-J_B)\cos k_x\cos k_y
-\frac{\Delta_{\mathrm{eff}}}{2}
\right]\sigma_z ,
\end{aligned}
\label{eq:bloch_hamiltonian}
\end{equation}
where $d_0(\mathbf{k})=2(J_A+J_B)\cos k_x\cos k_y$, $N_c$ is the number of unit cells, and $\sigma_{x,y,z}$ are the Pauli matrices.
The momentum components $k_x$ and $k_y$ are given in units of $1/a$.

From the Bloch Hamiltonian, we obtain the topological phase boundaries.
The phase boundaries are determined by the closing of the direct band gap.
In Eq.~\eqref{eq:bloch_hamiltonian}, this requires all three Pauli-matrix coefficients to vanish with $\sin k_x=\sin k_y=0$, and the phase boundaries are given by
\begin{equation}
\Delta_{c,\pm}=\pm4|J_A-J_B|,
\end{equation}
where $\Delta_{c,-}$ and $\Delta_{c,+}$ denote the left and right boundaries of the $C=1$ region, respectively.
Both boundaries are independent of $J_{AB}$.
Since $J_A$ and $J_B$ are independent of the layer separation $h$, the topological phase boundaries are independent of $h$.

For the model with cutoff $R_c=3a$, longer-range intrasublattice hoppings shift the topological phase boundaries, which remain independent of $h$, as shown in Fig.~2(a).

\section{Flux patterns and band topology of the coplanar configuration}
\label{sec:sm_bond_textures}

\begin{figure}[t]
\centering
\includegraphics[width=\columnwidth]{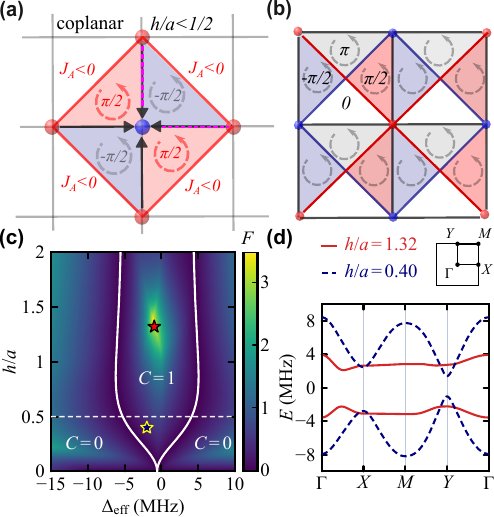}
\caption{
\textbf{Flux patterns and band topology of the coplanar configuration.}
(a) Hopping and staggered flux patterns around a site on sublattice $B$ for $0<h/a<1/2$.
(b) Representation of the staggered flux pattern using four effective fluxes.
Dashed arrows indicate counterclockwise circulation.
(c) Lower-band flatness $F$ as a function of $h/a$ and $\Delta_{\mathrm{eff}}$, with the topological phase boundaries marked by white solid lines.
The white dashed line marks $h/a=1/2$.
The set of bonds within $R_c=3a$ at $h/a=1.32$ is kept fixed throughout the scan.
(d) Band structures for uniform flux (red solid lines) at $(h/a,\Delta_{\mathrm{eff}}/\mathrm{MHz})=(1.32,-0.98)$ and staggered flux (blue dashed lines) at $(0.40,-2.00)$, corresponding to the two stars in (c).
Both lower bands have $C=1$.
}
\label{fig:flux_loops}
\end{figure}

We now consider the coplanar configuration with atomic positions $\mathbf r_{n,m}=(na,ma,(n+m)h)$, where sites with even and odd $n+m$ form sublattices $A$ and $B$, respectively.
The intrasublattice NN hoppings can be classified into two types, with displacement vectors $\mathbf r_1=(a,-a,0)$ and $\mathbf r_2=(a,a,2h)$.
The hopping amplitudes along $\mathbf r_1$ are independent of $h$, while for $\mathbf r_2$, the angular dependence is
\begin{equation}
1-3\cos^2\theta=\frac{a^2-4h^2}{a^2+2h^2},
\label{eq:sm_coplanar_sign}
\end{equation}
which changes sign at $h=a/2$.
The elementary triangles can be classified into two types, with intrasublattice bonds along $\mathbf r_1$ and $\mathbf r_2$, respectively.
For $h/a>1/2$, both types of triangles carry the same flux $\Phi=-\pi/2$, forming a uniform flux pattern.
For $0<h/a<1/2$, triangles with bonds along $\mathbf r_1$ enclose a flux of $-\pi/2$, while those with bonds along $\mathbf r_2$ enclose a flux of $+\pi/2$.
This gives rise to a staggered flux pattern, as shown in Figs.~\ref{fig:coplanar}(c) and \ref{fig:flux_loops}(a).

To better describe the staggered flux pattern, we introduce an effective flux pattern consisting of four fluxes, as shown in Fig.~\ref{fig:flux_loops}(b).
Each square is divided into four triangular regions with effective fluxes $(f_d,f_r,f_u,f_l)=(0,\pi/2,\pi,-\pi/2)$, corresponding to the lower, right, upper, and left regions, respectively.
The sum of the effective fluxes in two adjacent regions gives the flux through the corresponding elementary triangle, with $f_d+f_r=f_u+f_l=\pi/2$ and $f_d+f_l=f_r+f_u=-\pi/2$ modulo $2\pi$.
%

We further calculate the Chern number and flatness $F$ of the lower band as functions of $h/a$ and $\Delta_{\mathrm{eff}}$, as shown in Fig.~\ref{fig:flux_loops}(c).
We find that the lower band can have $C=1$ for both uniform and staggered flux.
Figure~\ref{fig:flux_loops}(d) shows representative band structures for the two cases.
For uniform flux, we choose $h/a=1.32$ and $\Delta_{\mathrm{eff}}=-0.98~\mathrm{MHz}$ with $F=3.27$ for the many-body study of the coplanar configuration.
The lower band shown for staggered flux is dispersive and carries a Chern number $C=1$.
A detailed study of the staggered flux pattern is beyond the scope of this work.

For the adiabatic interpolation in Eq.~\eqref{eq:interpolation} of the main text, we apply a local gauge transformation $\widetilde c_i=e^{i\varphi_i}c_i$ to the bilayer Hamiltonian.
For a site $i$ at $(x,y)=(na,ma)$, the phase $\varphi_i$ is given by
\begin{equation}
\varphi_i=
\begin{cases}
\frac{\pi}{2}(n+m), & i\in A,\\[4pt]
\frac{\pi}{2}(n+m-1), & i\in B.
\end{cases}
\label{eq:sm_gauge}
\end{equation}
The hopping amplitudes transform as $\widetilde t_{ij}=e^{i(\varphi_j-\varphi_i)}t_{ij}$.
This transformation matches the NN hopping phases of the bilayer array to those of the coplanar configuration without changing the flux through any closed loop.

\setcounter{topnumber}{4}
\setcounter{bottomnumber}{2}
\setcounter{totalnumber}{6}
\renewcommand{\topfraction}{0.95}
\renewcommand{\bottomfraction}{0.95}
\renewcommand{\textfraction}{0.05}

\section{System parameters and numerical details}
\label{sec:sm_numerical_details}

We use $a=9~\mu\mathrm{m}$, $B_z=46~\mathrm{G}$~\cite{Emperauger2025}, and a cutoff of $R_c=3a$ for the many-body calculations.
For the bilayer array, we choose $h/a=1.02$ and $\Delta_{\mathrm{eff}}=-2.77~\mathrm{MHz}$.
The density-density interactions $V_{ij}$ are calculated using PairInteraction~\cite{WeberPairInteraction2017}:
\begin{equation}
V_{ij}=U_{11}+U_{00}-U_{10}-U_{01},
\label{eq:sm_vij}
\end{equation}
where $U_{\alpha\beta}$ denotes the van der Waals shift of the two-atom state.
The hopping amplitudes $t_{ij}$ and density-density interactions $V_{ij}$ for the bilayer and coplanar configurations are listed in Tables~\ref{tab:microscopic} and \ref{tab:microscopic_coplanar}, respectively.

\begin{table}[!htbp]
\caption{
Hopping amplitudes and density-density interactions for the bilayer array.
For $AB$ bonds, $\mathbf r_{ij}$ is directed from sublattice $A$ to sublattice $B$, and only the hopping magnitudes are listed.
}
\label{tab:microscopic}
\centering
\begin{ruledtabular}
\begin{tabular}{cccc}
Bond & $\mathbf r_{ij}$ & $t_{ij}$ (MHz) & $V_{ij}$ (MHz)\\
\hline
$AA$ & $(\pm a,\pm a,0)$  & $-1.07$ & $-0.19$\\
$AA$ & $(\pm2a,0,0)$      & $-0.38$ & $-2.30\times10^{-2}$\\
$AA$ & $(0,\pm2a,0)$      & $-0.38$ & $-2.30\times10^{-2}$\\
$AA$ & $(\pm2a,\pm2a,0)$  & $-0.13$ & $-2.90\times10^{-3}$\\
$BB$ & $(\pm a,\pm a,0)$  & $1.27$  & $1.11\times10^{-3}$\\
$BB$ & $(\pm2a,0,0)$      & $0.45$  & $1.40\times10^{-4}$\\
$BB$ & $(0,\pm2a,0)$      & $0.45$  & $1.40\times10^{-4}$\\
$BB$ & $(\pm2a,\pm2a,0)$  & $0.16$  & $1.70\times10^{-5}$\\
$AB$ & $(\pm a,0,h)$      & $1.70$  & $-4.15\times10^{-2}$\\
$AB$ & $(0,\pm a,h)$      & $1.70$  & $-4.15\times10^{-2}$\\
$AB$ & $(\pm2a,\pm a,h)$  & $0.25$  & $-1.90\times10^{-3}$\\
$AB$ & $(\pm a,\pm2a,h)$  & $0.25$  & $-1.90\times10^{-3}$\\
\end{tabular}
\end{ruledtabular}
\end{table}

\begin{table}[!htbp]
\caption{
Hopping amplitudes and density-density interactions for the coplanar configuration.
For $AB$ bonds, $\mathbf r_{ij}$ is directed from sublattice $A$ to sublattice $B$, and only the hopping magnitudes are listed.
}
\label{tab:microscopic_coplanar}
\centering
\begin{ruledtabular}
\begin{tabular}{cccc}
Bond & $\mathbf r_{ij}$ & $t_{ij}$ (MHz) & $V_{ij}$ (MHz)\\
\hline
$AA$ & $\pm(a,-a,0)$    & $-1.07$ & $-0.19$\\
$AA$ & $\pm(a,a,2h)$    & $0.15$  & $-5.50\times10^{-4}$\\
$AA$ & $\pm(2a,-2a,0)$ & $-0.13$ & $-2.90\times10^{-3}$\\
$BB$ & $\pm(a,-a,0)$    & $1.27$  & $1.11\times10^{-3}$\\
$BB$ & $\pm(a,a,2h)$    & $-0.18$ & $-2.73\times10^{-4}$\\
$BB$ & $\pm(2a,-2a,0)$  & $0.16$  & $1.70\times10^{-5}$\\
$AB$ & $\pm(a,0,h)$     & $1.05$  & $-2.33\times10^{-2}$\\
$AB$ & $\pm(0,a,h)$     & $1.05$  & $-2.33\times10^{-2}$\\
$AB$ & $\pm(2a,-a,h)$   & $0.25$  & $-1.09\times10^{-3}$\\
$AB$ & $\pm(-a,2a,h)$   & $0.25$  & $-1.09\times10^{-3}$\\
\end{tabular}
\end{ruledtabular}
\end{table}

\begin{figure}[!htbp]
\centering
\includegraphics[width=\columnwidth]{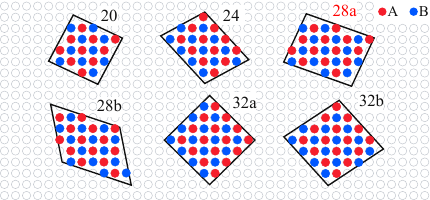}
\caption{
\textbf{Torus geometries for exact diagonalization.}
The numbers denote the number of sites $N_s$, and labels $a$ and $b$ distinguish different geometries with the same $N_s$.
Black outlines indicate the boundaries of the tori.
}
\label{fig:torus_shapes}
\end{figure}

We perform exact diagonalization on tori of different sizes and shapes, as shown in Fig.~\ref{fig:torus_shapes}.
%
%
To characterize the many-body topology, we impose twisted boundary conditions along the two translation vectors $\mathbf L_1$ and $\mathbf L_2$ of each torus,
\begin{equation}
\Psi(\mathbf r+\mathbf L_\mu)
=e^{i\theta_\mu}\Psi(\mathbf r),
\qquad \mu=1,2.
\label{eq:sm_twisted_boundary}
\end{equation}
All the hopping terms crossing the periodic boundaries acquire the corresponding phase factors.

\begin{figure}[!htbp]
\centering
\includegraphics[width=\columnwidth]{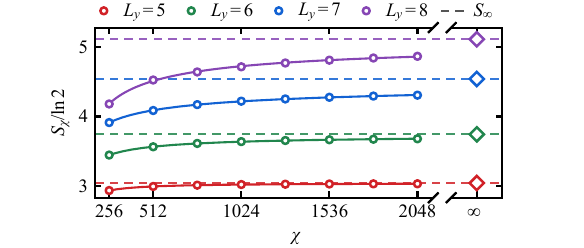}
\caption{
\textbf{Entanglement entropy extrapolation.}
Solid curves show fits to Eq.~\eqref{eq:sm_entropy_extrapolation}.
Horizontal dashed lines and diamonds indicate the extrapolated entanglement entropies at $\chi\to\infty$.
}
\label{fig:tee_extrapolation}
\end{figure}

For the entanglement entropy calculations, we consider infinite cylinders with $L_y=5,6,7,8$.
For each $L_y$, we extrapolate the entanglement entropy to infinite bond dimension using~\cite{Weber2022}
\begin{equation}
S_\chi=S_\infty-A\chi^{-p},
\label{eq:sm_entropy_extrapolation}
\end{equation}
with bond dimensions $256\leq\chi\leq2048$, as shown in Fig.~\ref{fig:tee_extrapolation}.
We then fit the extrapolated entropies to $S_\infty(L_y)=\alpha L_y-\gamma$, obtaining $\gamma\simeq0.44\ln2$.

\section{Many-body spectral flow in the coplanar configuration}
\label{sec:sm_coplanar_spectral_flow}

\begin{figure}[!htbp]
\centering
\includegraphics[width=\columnwidth]{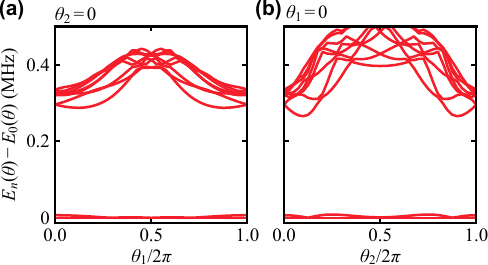}
\caption{
\textbf{Many-body spectral flow in the coplanar configuration.}
(a) Varying $\theta_1$ from $0$ to $2\pi$ at fixed $\theta_2=0$.
(b) Varying $\theta_2$ from $0$ to $2\pi$ at fixed $\theta_1=0$.
Energies are measured relative to $E_0(\theta_1,\theta_2)$.
}
\label{fig:coplanar_spectral_flow}
\end{figure}

We further examine the low-energy spectrum of the coplanar configuration on the $N_s=28$ torus under flux insertion along both periodic directions, as shown in Fig.~\ref{fig:coplanar_spectral_flow}.
The two lowest states remain separated from higher excitations by a finite many-body gap.

\section{Dependence on the hopping range}
\label{sec:sm_cutoff_dependence}

To examine the effect of longer-range dipolar hoppings, we optimize $h/a$ and $\Delta_{\mathrm{eff}}$ for each cutoff $R_c$ by maximizing the lower-band flatness $F$ within the $C=1$ region.
The optimal parameters and the critical detunings $\Delta_{c,\pm}$ for the bilayer and coplanar configurations are listed in Tables~\ref{tab:cutoff_dependence} and \ref{tab:cutoff_dependence_coplanar}, respectively.
For both cases, a finite $C=1$ region persists for all cutoffs considered.
Even at $R_c=100a$, the lower-band flatness reaches $F=3.75$ and $F=2.69$ for the bilayer and coplanar configurations, respectively.

\begin{table}[!ht]
\caption{
Optimal parameters and lower-band flatness for different hopping cutoffs in the bilayer array.
NN denotes the model with intersublattice and intrasublattice NN hoppings.
All detunings are in MHz.
}
\label{tab:cutoff_dependence}
\centering
\begin{ruledtabular}
\begin{tabular}{crrrrr}
$R_c$ & $\Delta_{c,-}$ & $\Delta_{c,+}$ & $h/a$ & $\Delta_{\mathrm{eff}}$ & $F$\\
\hline
NN     & $-9.38$  & $9.38$ & $1.01$ & $0.49$  & $4.78$\\
$3a$   & $-13.86$ & $4.89$ & $1.02$ & $-2.77$ & $8.53$\\
$4a$   & $-15.95$ & $6.15$ & $0.92$ & $-3.32$ & $8.01$\\
$5a$   & $-16.89$ & $5.91$ & $0.88$ & $-3.26$ & $5.99$\\
$10a$  & $-19.12$ & $6.20$ & $0.85$ & $-4.39$ & $4.57$\\
$20a$  & $-20.14$ & $6.20$ & $0.81$ & $-4.67$ & $4.10$\\
$50a$  & $-20.76$ & $6.20$ & $0.78$ & $-4.81$ & $3.83$\\
$100a$ & $-20.97$ & $6.20$ & $0.77$ & $-4.86$ & $3.75$\\
\end{tabular}
\end{ruledtabular}
\end{table}

\begin{table}[!ht]
\caption{
Optimal parameters and lower-band flatness for different hopping cutoffs in the coplanar configuration.
The critical detunings $\Delta_{c,\pm}$ bound the $C=1$ region at the optimal $h/a$ for each cutoff.
The symbols $\dagger$ and $\ddagger$ indicate that the optimal values of $h/a$ approach $\sqrt{7}/2$ from below and $4/3$ from above, respectively.
}
\label{tab:cutoff_dependence_coplanar}
\centering
\begin{ruledtabular}
\begin{tabular}{crrrrr}
$R_c$ & $\Delta_{c,-}$ & $\Delta_{c,+}$ & $h/a$ & $\Delta_{\mathrm{eff}}$ & $F$\\
\hline
NN     & $-5.42$ & $5.42$ & $1.22$            & $0.40$  & $2.48$\\
$3a$   & $-5.93$ & $4.76$ & $1.32^\dagger$   & $-0.98$ & $3.27$\\
$4a$   & $-6.67$ & $4.16$ & $1.23$            & $-1.04$ & $3.57$\\
$5a$   & $-6.85$ & $4.36$ & $1.33^\ddagger$  & $-1.37$ & $3.38$\\
$10a$  & $-6.59$ & $4.32$ & $1.35$            & $-1.51$ & $3.00$\\
$20a$  & $-6.57$ & $4.30$ & $1.38$            & $-1.59$ & $2.82$\\
$50a$  & $-6.57$ & $4.30$ & $1.40$            & $-1.63$ & $2.72$\\
$100a$ & $-6.56$ & $4.30$ & $1.41$            & $-1.65$ & $2.69$\\
\end{tabular}
\end{ruledtabular}
\end{table}

\end{document}